\documentclass[12pt]{iopart}
\usepackage{amsfonts}
\usepackage{amsmath}
\usepackage{booktabs}
\usepackage{siunitx}
\usepackage[justification=centering]{caption}
\usepackage{hyperref}
\usepackage{arydshln}
\usepackage{pdfpages}
\usepackage[style=phys,sorting=none,maxbibnames=6]{biblatex}
\hypersetup{
    colorlinks=true,
    linkcolor=blue,
    citecolor=blue,
    filecolor=blue,      
    urlcolor=blue
    }

\begin{document}
\title[]{Comparison of arm cavity optical losses for the two wavelengths of the Einstein Telescope gravitational wave detector}

\author{Maxime Le Jean$^{1}$, Jerome Degallaix$^1$, David Hofman$^1$, Laurent Pinard$^1$, Danièle Forest$^1$, Massimo Granata$^1$, Christophe Michel$^1$, Jessica Steinlechner$^{2,3}$, Claude Amra$^4$, Michel Lequime$^4$ and Myriam Zerrad$^4$}

\address{
1.\textit{ Laboratoire des Matériaux Avancés - IP2I, CNRS/IN2P3, Université de Lyon 69100 Villeurbanne}\\

2.\textit{ Maastricht University, Minderbroedersberg 4-6, 6211 LK Maastricht, The Netherlands}\\

3.\textit{ Nikhef, Science Park 105, 1098 XG Amsterdam, The Netherlands}\\

4.\textit{ Aix Marseille Univ, CNRS, Centrale Med, Institut Fresnel, Marseille, France}
}

\ead{lejean@lma.in2p3.fr}
\vspace{10pt}
\begin{indented}
\item[]January 2024
\end{indented}

\begin{abstract}

A new generation of gravitational wave detectors is currently being designed with the likely use of a different laser wavelength compared to current instruments.
The estimation of the optical losses for this new wavelength is particularly relevant to derive the detector sensitivity and also to anticipate the optical performances of future instruments. 
In this article, we measured the absorption and angle-resolved scattering of several mirror samples in order to compare optical losses at a wavelength of 1064 and 1550\ nm.
In addition, we have carried out simulations of the Einstein Telescope arm cavities at 1064 and 1550\ nm taking into account losses due to surface low-spatial frequency flatness. Our results suggest that optical losses as measured at 1064\ nm are about twice as large as those at 1550\ nm as predicted with a simple model.

\end{abstract}

\section{The Einstein Telescope}

Direct detections of Gravitational Waves (GW) on Earth have shed light on the Universe's most cataclysmic events \cite{GW_detection}. Very compact objects such as black holes or neutron stars are merging, sending ripples in the space-time sea.
The waveforms of those ripples, as recorded with giant laser interferometers, give unprecedented information on the mass, spin, deformability, or distance of their progenitors \cite{Coalescence_parameters}.

While the collaborations LIGO, Virgo, and KAGRA, operating the eponymous GW detectors have claimed numerous discoveries \cite{abbott2023population} and are still recording data \cite{abbott2020prospects}, the next generation of instruments is currently under design.
These new infrastructures, such as the Einstein Telescope (ET) in
Europe \cite{ET} or Cosmic Explorer (CE) in the USA \cite{CE} aim to be ten times more sensitive than current detectors, targeting the majority of black hole mergers in the Universe with a horizon reaching a redshift up to one hundred \cite{maggiore2020science}.

In particular, the Einstein Telescope detector, which is expected to commence operation around 2035, is very ambitious.
Not only the sensitivity will be increased by ten fold, but another objective is to widen the detector's bandwidth, in particular by extending it to low frequencies, with the detection range starting from a few Hertz \cite{ET}. 
Such an impressive feat will be achieved thanks to a unique strategy, as ET will be composed of two detectors, one dedicated to low frequency (ET-LF), and one for high frequency (ET-HF) and the data from both instruments will be recombined.
Those two detectors will share a common length, shape, and orientation but their main components and operating temperature will differ.

While ET-HF is based on the same technology as LIGO and Virgo, albeit with specifications pushed to the extreme, ET-LF \cite{ET} will be radically different and with a technological approach similar to the KAGRA detector \cite{abe2022KAGRA}.
The mirrors of the long arm cavities will be cooled down to the temperature of 10-20\ K and made of silicon with the laser wavelength shifted from 1064\ nm (similar to current detectors) to 1550\ nm.

The use of a longer laser wavelength in the interferometer is mandatory to benefit from the transparency range of the silicon substrate of the arm cavity mirrors.
Another advantage is the reduction of the optical loss due to the decrease in the amount of light scattered from surface imperfections \cite{Stover}.
Light scattering is the dominant source of optical loss in the long arm cavities with the detrimental effect of lowering the power recycling gain as well as inducing extra phase noise if the scattered light happens to recombine with the main laser beam after being reflected on vibrating elements \cite{PhaseNoise}. Moreover, optical losses could also directly degrade the detector sensitivity by limiting the level of frequency-dependent squeezing, since those losses could be amplified inside the filtering cavity \cite{Squeezing}.

At first approximation, the reduction in scattering loss is directly proportional to the ratio of the wavelength squared \cite{Stover}.
So by shifting the laser wavelength from 1064\ nm to 1550\ nm, a reduction of 50\% could be expected.

In this article, we report the first exhaustive comparative study regarding the optical loss at the two relevant wavelengths for the next generation of GW detectors.
We produced dedicated high reflectivity mirrors for both wavelengths of ET, made with materials representative of those used in GW detectors \cite{Pinard:17},\cite{granata2020amorphous}. 
The study includes the measurement of optical absorption and optical scattering, an assessment of the point-like defect content, and a complementary analysis based on optical simulations to reveal the impact of low-frequency surface errors for ET-HF and ET-LF.

\section{Mirror manufacturing}

For this study, we manufactured two types of mirrors, one with a highly reflective coating centered at 1064 nm and the other centered at 1550\ nm. 
Each type of reflective coating was deposited on 1" and 2" diameter micro-polished fused silica substrates from Coastline Optics with a roughness specification under 1\ \r{A} RMS and a flatness of $\lambda$/20 P-V at 633\ nm in the 80\% clear aperture.
The coatings consist of a quarter-wave multi-layer stacks alternating a high-index material (Ta$_2$O$_5$: $n=2.04$) and a low-index material (SiO$_2$: $n=1.45$) centered on the respective wavelength.
The coatings were produced using a commercial VEECO ion-beam sputtering (IBS) deposition system in a vacuum chamber using argon as sputtering ions. This deposition chamber produces coatings with a thickness uniformity deviation of 0.06\% P-V over 80\ mm diameter. 

For comparison purposes, the two kinds of coating have the same design and are made of 34 layers.
The transmission of both is the same and in the order of 40\ ppm.
The coating total thickness for the mirror at 1064\ nm is 5.13\ µm whereas the one for 1550\ nm is 7.68\ µm. 
The choice of material was driven by the knowledge of their properties in terms of refractive index, low optical loss, and low thermal noise, which are of great interest for any optical precision experiments \cite{granata2020amorphous}. 
A total of 8 different samples of 1 or 2 inches for both 1064\ nm and 1550\ nm wavelength have been produced.
All samples for a given wavelength were coated in the same run. 
We can therefore assume that the properties imparted to these mirrors by the coatings are identical since from the exact same process.  Even if the samples are not both at the same position in the deposition chamber, the coating uniformity should guarantee this hypothesis.
Individual characterisation was done after coating and annealing the mirrors at 500°C for 10 hours with ramps of 100°C per hour which is the procedure adopted for the current generation of gravitational-wave detector mirrors.

However we have subsequently observed great variability in the optical loss measurements detailed in this article, likely linked to defects already present on the uncoated substrates.
To ease the comparison, we will discuss only the best samples regarding the optical losses at the two different wavelengths.

\section{Scattering losses comparison}

To a first approximation, the reduction in scattering loss is directly proportional to the wavelength squared ratio \cite{Stover}. 
Thus, excluding defects, we can expect a reduction of almost 50\% in the power of light scattered between 1064\ nm and 1550\ nm.
However, this approach may be too simplistic as high reflectivity coatings require to be thicker for longer wavelength, which could imply a higher defect density \cite{Sihem}, mitigating the expected decrease in optical loss.

\subsection{Scattering measurement setup}

Light scattering characterization was performed at the Institut Fresnel using a custom-built Spectral and Angular Light Scattering Apparatus (SALSA), managed by the Light Scattering Group of the Institut Fresnel \cite{SALSA}.
SALSA is based on a supercontinuum laser source emitting from 400\ nm to 2400\ nm, which is split by a dichroic filter into two different channels, a visible channel (400-905\ nm) and a near infra-red channel (905-1650\ nm).
Each of these channels is equipped with a tunable volume hologram filter, which allows isolating in these broadband spectra a narrow line whose center wavelength can be continuously tuned over the corresponding channel wavelength range.
A grating spectrograph is inserted just before the detection assembly (2D CCD array for the visible channel,
InGaAs linear array for the near-infrared channel) to improve the spectral resolution of the measurement (typically 1\ nm) and to minimize the level of parasitic light.
This photometric instrument can be used either to measure the spectral dependence of the transmittance of a flat sample up to an optical density of 13 or to record the angle-resolved scattering (ARS) of the same sample with a detection floor of about 10$^{-8}$\ sr$^{-1}$ and with an angular resolution better than 1$^\circ$.
The two-inch samples were characterized at 1064\ nm and 1550\ nm with SALSA, allowing their scattering losses to be determined by calculating the Total Integrated Scattering (TIS) through the integration of the Angle Resolved Scattering ARS($\theta$) :

\begin{equation}
\label{TIS}
\centering
TIS = 2\pi \displaystyle\int_{\theta_i+1^\circ}^{\frac{\pi}{2}} ARS(\theta)sin(\theta)d\theta
\end{equation}

where $\theta_i$ is the angle of incidence of the light beam on the sample (here 5 degrees).

\subsection{Results}

A total of 6 samples were measured, including 2 blank substrates.
As explained earlier, the results showed some variability, even between the two different blank substrates.
Regarding the uncoated substrates, the best one has a total TIS of 5 ppm at 1550\ nm.
This measurement includes the contributions of both the front and the back sides.
Thanks to a numerical model \cite{Amra}, assuming both sides to be identical, the TIS per side is estimated to be 2\ ppm.

\begin{figure}[!]
    \centering
        \includegraphics[clip,width=0.65\textwidth]{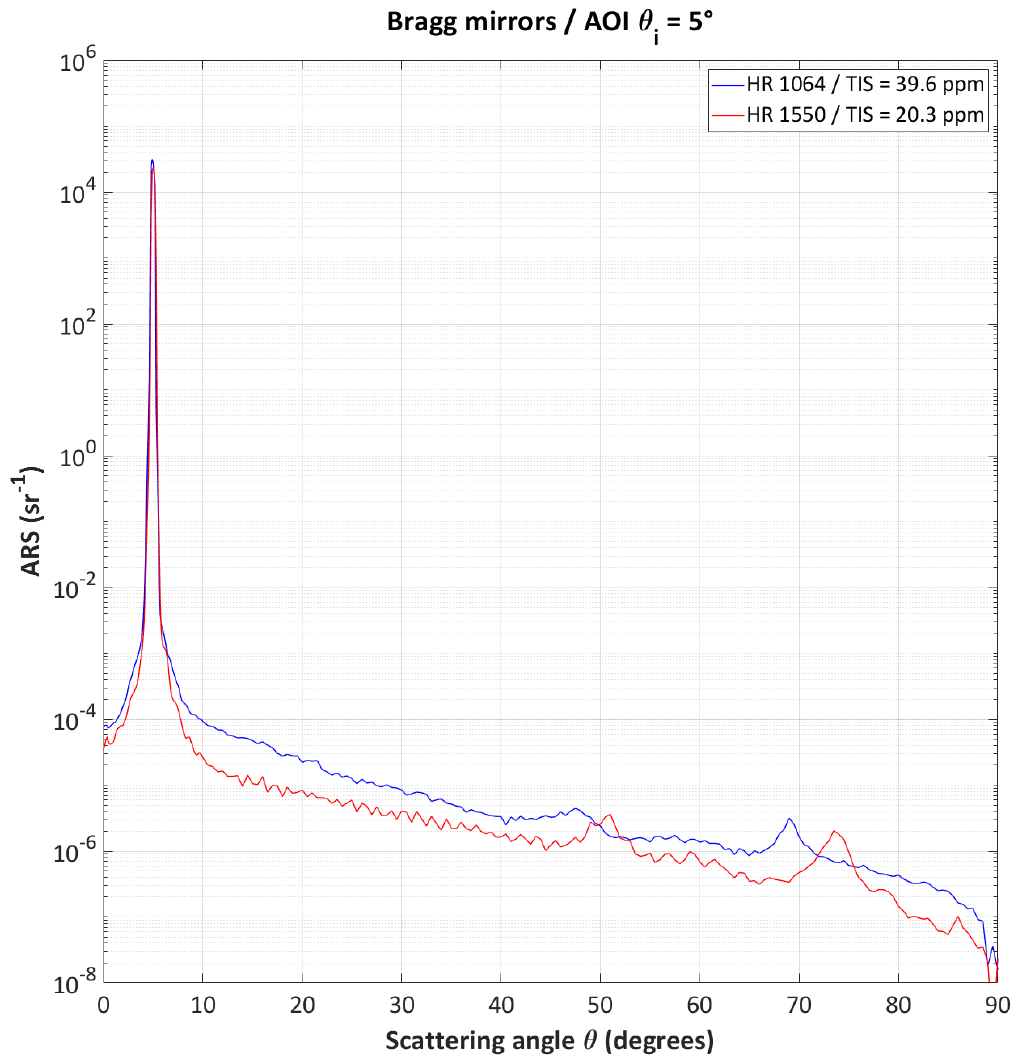}
    \caption{Measured Angle Resolved Scattering (ARS). The blue (respectively red) curve shows the ARS of the best high reflectivity coating at 1064\ nm (respectively 1550\ nm).}
    \label{fig:BRDF}
\end{figure}

For the coated mirrors, the total scattering of the best sample at 1064\ nm was measured to be 39.6\ ppm, while the scattering of the best sample at 1550\ nm was measured to be 20.3\ ppm see fig.(\ref{fig:BRDF}). 
As expected the scattering losses are significantly lower for the 1550\ nm mirrors by a factor of nearly 2.

\section{Coating optical absorption}
The coating optical absorption is also a critical parameter that could lead to tragic consequences in the operation of the detector. The issue is not from the extra optical loss as such (usually negligible) but rather from the thermal wavefront distortions induced by the combination of very high laser power and optical absorption.
As an example, for current GW room temperature detectors, several hundreds of kilowatts of light are circulating in the arm cavities, requiring a complex system of thermal compensation \cite{rocchi2012thermal} even though the coating absorption is below 1\ ppm (parts per million, 10$^{-6}$).
For ET, in the high-frequency interferometer, 3\ MW of light is expected to be incident on the large mirrors.

For cryogenic interferometers, the problem is different as the wavefront distortions due to the optical absorption become negligible thanks to the high thermal conductivity, lower thermo-refractive, and expansion coefficients of the substrate at low temperatures.
However, the issue will come from the amount of heat absorbed in the coating and substrate, which is preventing the efficient cooling of the mirrors. 

To measure the optical absorption on our samples, the photo-deflection technique (also called the mirage effect) is used \cite{jackson_photothermal_nodate}.
A high-power pump laser at the wavelength of interest (1064\ nm or 1550\ nm) is focused on the HR coating of the sample.
A small part of the light is absorbed within the coating, generating a hot spot and hence a gradient of temperature within the substrate. A second low-power laser at a different wavelength (1310\ nm in our case) is deflected by this gradient and the displacement is recorded on a quadrant photodiode.
After comparison with a reference in absorption, this probe beam displacement enables the derivation of the absolute optical absorption (given with an accuracy of 10\%).

The absorption of the mirror at 1064\ nm was measured at 0.7\ ppm (with a pump laser at 1064\ nm) whereas the absorption of the mirror at 1550\ nm was found to be 10 times higher at 6.7\ ppm (with a pump laser at 1550\ nm).
The relatively high absorption at 1550\ nm was not expected, so we decided to cross-check this result independently.

One of our two 1550\ nm samples was therefore sent to Maastricht University for its optical absorption to be characterized by a Photo-thermal Common path Interferometry setup \cite{PCI}. 
The optical absorption at 1550 nm was measured at 7.8 ppm which confirms the high absorption of the coating stack for this wavelength.
 
To be noted that for current gravitational wave detectors operating at room temperature, the high refractive index layers are not made of pure tantala but rather tantala doped with titania. The benefits of this mixture are coatings with lower mechanical losses \cite{granata2020amorphous} and twice lower optical absorption.
For ET-HF and ET-LF, the exact coating formula has yet to be determined, as intense worldwide research effort is underway to find better materials \cite{Granata:20}. 
So the absorption values reported here are mentioned for completeness but preclude any conclusion regarding the Einstein Telescope since the choice of coating material is not decided yet.

\section{Point-like defects}

Point defects in the coating layers could also account for a fraction of scattering losses.
We characterized each sample after annealing using a customly made by EOTECH optical profilometer to detect the point defects in dark field conditions.
The instrument produces cartography of defects made with 569 images of 1024x1024 pixels with a pixel resolution of 0.45\ µm² covering a total disk of 18\ mm diameter.
All images are taken in the same illumination condition.
The instrument is capable of detecting point-like defects above 1 micron in diameter.
After the dark field images has been recorded by a CCD camera, point defect detection is performed with an automatic algorithm that allowing us to count the number of defects and extract the surface area of each point defect detected.
Previous studies done over monolayer coatings have shown that the defects present in the coating can be of different kind: bubbles, particulate contaminants, or cracks. \cite{Sihem}.

\setlength{\tabcolsep}{15pt}
\renewcommand{\arraystretch}{1.2}
\begin{table}[]
    \caption{Summary of detected defects giving the count of defects, the average area of the defects, and the defect density to the surface area.}
\begin{center}
    \begin{tabular}{cccc} \toprule
    {Mirror's Wavelength} & {Counts}  & {$\langle$Area$\rangle$ [µm$^{2}$]}  & {Density [mm$^{-2}$]}     \\ \midrule
    1064  & 4426 & 7.27  & 13.66   \\
    \\[-2.6ex] \hdashline\\[-6.6ex]\\
    1550  & 2411 & 11.15 & 7.44   \\ \bottomrule
    \end{tabular}

    \label{tab:defects}
\end{center}
\end{table}

For these multilayer coatings on the 1 inch samples, we found out that the total number of detected defects was not proportional to the coating thickness, as shown in Table \ref{tab:defects}.
However, our results suggest that the defect size is proportional to the thickness of the coating.
This leads mirrors at 1550\ nm to have larger defects, making them more susceptible to induce light scattering. 

\section{Impact of low spatial frequency surface errors}

Since arm cavity mirrors are of finite size, the low-spatial-frequency fluctuations in surface height lead to low-angle scattering, which sends light out of the cavity.
Any deviation from the perfect spherical shape of the mirrors can also induce coupling loss as the resonant mode of the cavity is no longer a perfect Gaussian fundamental mode matched to the input beam.
These latter kinds of losses are not considered in the following simulations as only clipping loss will be reported. 

\subsection{Numerical implementation with surface maps}
To study these effects, we have carried out optical cavity simulations taking into account arbitrary surface profiles of the mirror. 
One particularity of the simulations for the Einstein Telescope is that the mirrors will be twice as large as the cavity mirrors of the second generation of gravitational wave detectors.  Therefore, the surface spatial fluctuation spectrum extends into a low-frequency range not sampled by the current LIGO/Virgo mirrors. 
It is then necessary to extrapolate the spatial fluctuation spectrum.
We have applied a procedure based on Power Spectral Density (PSD) that will allow us to extrapolate the surface properties of mirrors measured in the laboratory, into simulated surface height maps of arbitrary size. 
The PSD in 2 dimensions is defined as follows :   
\begin{equation}
\label{eq:PSD}
\centering
    PSD^{2D}_{q_x,q_y} = \frac{(\Delta x\Delta y)^2}{A}\left|\tilde h_{x,y} \right|^2  
\end{equation}
\noindent
where $\tilde h_{x,y}$ is the 2D Fourier transform of the discrete surface height deviation, sampled with spatial resolution, $\Delta x$, $\Delta y$, $A$ the area of the surface probed and $q_x$, $q_y$ the spatial frequency of the height fluctuation along the $x$ and $y$ axis.

The PSD as described in Eq.(\ref{eq:PSD}) decomposes the height of a surface into the contribution of a set of harmonics over a spectrum of spatial frequencies. 
To derive a representative PSD$^{2D}$, we measured the height $h_{x,y}$ of 7 large arm cavity mirrors which were polished for the Advanced Virgo interferometer.
Those mirrors were polished using an ion beam figuring method \cite{rauschenbach2022ion} and represent the state-of-the-art performances for large optics.
The flatness measurements were achieved with a customized Zygo wavelength shifting interferometer. This Fizeau interferometer has sub-nanometer height accuracy over a surface area of 300\ mm diameter.
We then derived the PSD$^{1D}$ in one dimension using an azimuthal integration, i.e. at a constant frequency.

\begin{figure}[htbp]
    \centering
        \includegraphics[clip, trim=0.5cm 9cm 0.5cm 9.5cm, width=0.90\textwidth]{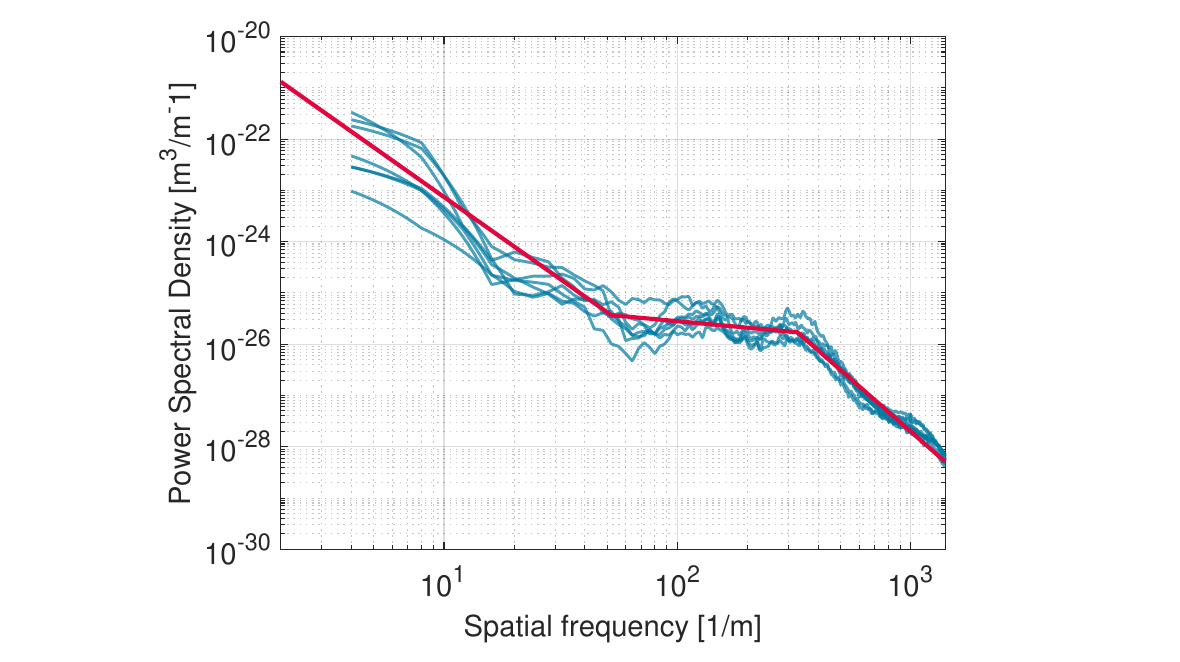}
    \caption{Blue curves are the computed  PSD$^{1D}$ of the 7 measured mirrors surface height. The red curve is the result of our fit with the extrapolation toward the lower spatial frequency range.}
    \label{fig:PSD_FIT}
\end{figure}

The PSD$^{1D}$ is then fitted over three continuous segments, each behaving as a simple power law.
We can then extrapolate this one-dimensional PSD fit to the low spatial frequency range to simulate mirrors of a bigger size.
After the extrapolation step, the PSD$^{1D}$ is converted back into two dimensions by assuming radial symmetry along the frequencies, which is equivalent to considering spatial anisotropy of the mirror surface.
A pixel-by-pixel random phase is then added to the complex PSD$^{2D}$ before the inverse transformation to the real height space.

\subsection{Cavity Simulations}

We used the OSCAR optical simulation package \cite{OSCAR} to perform optical simulations of the arm cavities of the Einstein Telescope.
In OSCAR, the laser beam propagation is performed using a Fast Fourier Transform (FFT) algorithm on a 2D grid sampling the electromagnetic field of the light.
OSCAR can take into account the arbitrary surface height of the mirrors, allowing the calculation of the optical scattering loss inside the cavities.

We used the physical mirror diameters specified in the technical design report, \cite{ET_Design} as references and investigated the remaining parameters for ET cavities in greater detail.
Note that all our consideration is based on symmetrical cavities of length $L = 10\ $km.

To ensure comparable arm cavity simulations between 1064\ nm and 1550\ nm, we have chosen the radii of curvature so that the ratio of the mirror diameter $d_m$ to the diameter of the beam on the mirror is similar. 
This led us to take radius of curvature values as RoC$_{LF}$=5950\ m and RoC$_{HF}$=5090\ m yielding a mirror-to-beam diameter ratio of 2.75 for both ET-LF and ET-HF. 
To validate this value of the radius of curvature, additional simulations were done to confirm that no higher-order optical modes were resonating at the same time as the fundamental mode for this geometry.
The non-degeneracy of the cavity guarantees the lowest optical losses.
The RoC chosen for our simulations are slightly different from the official ET ones, this is done on purpose to have a more meaningful comparison of the cavity losses between the 2 wavelengths.
In particular, in the case of perfect mirrors, the 2 cavities will have similar (negligible) optical losses since the mirror-to-beam diameter ratio is identical.

The optical simulations will calculate the steady state electric fields in the cavity and from there, we can derive the Round Trip Loss (RTL) as \cite{Straniero} : 

\begin{equation}
\label{RTL}
\centering
    RTL = \frac{P_{in}-P_{ref}-P_{trans}}{P_{circ}} 
\end{equation}

\noindent
where $P_{in}$ is the input power entering the cavity, $P_{ref}$ and $P_{trans}$ the reflected and transmitted power and $P_{circ}$ the power circulating in the cavity on resonance. 
We can define a central diameter by applying the surface irregularities only to a limited diameter of the mirror and study the induced loss. 
We were able to show that 98\% of the losses per cavity round trip were induced by the central 32\ cm diameter in the case of ET-HF and by the central 25\ cm diameter in the case of ET-LF.

In the following part, the surface flatness RMS of the mirror will be calculated over those central diameters $d_c$.
The simulation cavity parameters are summarized in the Table.\ref{tab:ET_Cavity}.

\setlength{\tabcolsep}{15pt}
\renewcommand{\arraystretch}{1.2}
\begin{table}[]
    \caption{Summary of the parameters used for ET cavity simulations.}
\begin{center}
    \begin{tabular}{cccccc} \toprule
    {Arm Cavity} & {$\lambda$ [nm]}  & {$d_m$ [m]}  & {$d_c$ [m]} & {RoC [m]} & {L [m]}    \\ \midrule
    ET-HF  & 1064 & 0.62 & 0.32 & 5090 & 10000  \\
    ET-LF  & 1550 & 0.45 & 0.25 & 5950 & 10000   \\ \bottomrule
    \end{tabular}
    \label{tab:ET_Cavity}
\end{center}
\end{table}

\begin{figure}[!]
    \centering
        \includegraphics[clip, trim=0.5cm 9cm 0.5cm 9.5cm, width=1.00\textwidth]{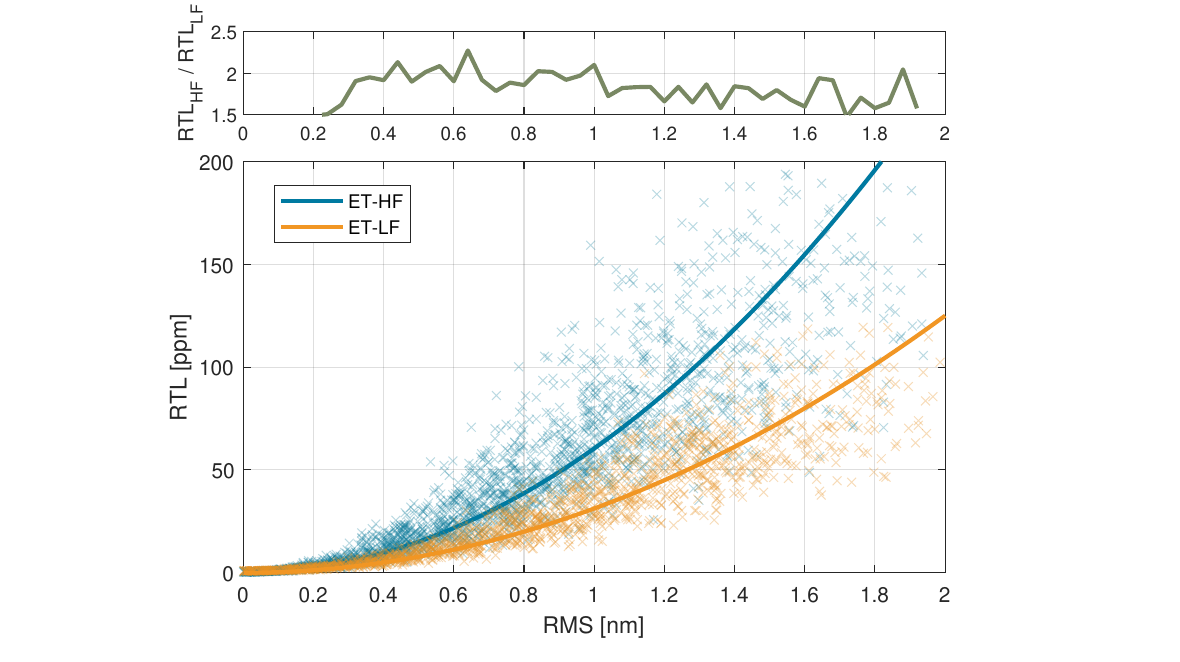}
    \caption{Bottom curves represent RTL as a function of the mirror flatness. In blue, is the ET-HF arm cavity (1064\ nm), and in orange is the ET-LF arm cavity (1550\ nm). Each point is the result of a cavity round-trip simulation. The top curve is the RTL ratio as a function of flatness level.} 
    \label{fig:RMS_RTL}
\end{figure}

\noindent

\subsection{Results} 
We have simulated the cavity optical round trip loss of the resonant fundamental beam for the ET arm cavity at both wavelengths with parameters shown in the Table \ref{tab:ET_Cavity}. 

More than 1500 cavities for both wavelengths were simulated with the flatness of the simulated mirrors randomly selected between 0 and 2\ nm over the central part.
The simulations' results are displayed in figure \ref{fig:RMS_RTL}.

According to the TIS law, which states that the TIS as computed with Eq.(\ref{TIS}) is proportional to $\sigma_{rms}^2$, we fitted the RTL of the 2 cavities with a quadratic law $RTL = a\sigma_{rms}^2$, represented by the continuous lines in figure \ref{fig:RMS_RTL}. 

The top graph in figure \ref{fig:RMS_RTL} is obtained by averaging the optical loss in bins of 0.02 nm RMS width and computing the ratio of the optical losses in the 2 cavities. We found that the RTL is about twice as low for the 1550\ nm arm cavity, compared to the 1064\ nm one, for the same mirror surface quality.

The results of these simulations already hint at some specifications for the flatness of the mirrors for ET.
Using a flatness RMS below 0.5\ nm, as achieved for the current laser, interferometers would also be compatible with ET.
The difficulty would lay in the manufacturing process, as the mirrors will be much heavier for ET and the central area with the most critical specification would be larger.

\section{Conclusion}

Our study confirms through experimental measurements and numerical simulations that the optical losses from scattering, induced by the mirrors inside resonant cavities are about a factor of two lower for a wavelength of 1550\ nm, compared to 1064\ nm.
The simple scattering model is validated despite having a thicker coating for 1550\ nm optics.
This work is intended to help the estimation of the optical performances of the future Einstein Telescope in the most realistic way.
However, once the design for the arm cavity is fixed, the estimation of the optical losses will be re-computed using the script developed for this article.

\section*{Acknowledgement}
Measurements performed at Maastricht University were part of the project No.VI.Vidi.203.062 financed by the Dutch Research Council (NWO).

\printbibliography
\end{document}